\documentclass[usenatbib]{mn2e}
\input{psfig}
\def\beq{\begin{equation}}
\def\eeq{\end{equation}}
\def\barray{\begin{eqnarray}}
\def\earray{\end{eqnarray}}

\newcommand{\Lsun}{\>{\rm L_{\odot}}}

\newcommand{\Msunh}{\>h^{-1}\rm M_\odot}

\def\gtsima{$\; \buildrel > \over \sim \;$}
\def\ltsima{$\; \buildrel < \over \sim \;$}
\def\prosima{$\; \buildrel \propto \over \sim \;$}
\def\gsim{\lower.7ex\hbox{\gtsima}}
\def\lsim{\lower.7ex\hbox{\ltsima}}
\def\simgt{\lower.7ex\hbox{\gtsima}}
\def\simlt{\lower.7ex\hbox{\ltsima}}
\def\simpr{\lower.7ex\hbox{\prosima}}
\def\la{\lsim}
\def\ga{\gsim}

\newcommand{\apj}{ApJ}
\newcommand{\apjs}{ApJS}
\newcommand{\apjl}{ApJL}

\newcommand{\mnras}{MNRAS}

\newdimen\hssize
\newdimen\hdsize
\def\lesssim{\mathrel{\hbox{\rlap{\hbox{\lower4pt\hbox{$\sim$}}}\hbox{$<$}}}}
\def\gtrsim{\mathrel{\hbox{\rlap{\hbox{\lower4pt\hbox{$\sim$}}}\hbox{$>$}}}}
\begin{document}


\title[The Local Reionization History]  
      {Dependence of the Local 
      Reionization History on Halo Mass and Environment: Did Virgo Reionize the Local Group?}
\author[S.M. Weinmann et al.]          
       {\parbox[t]{\textwidth}{
        Simone M. Weinmann$^{1}$\thanks{E-mail:weinmann@physik.uzh.ch},  
         Andrea V. Macci\`o$^{1,2}$, Ilian T. Iliev$^{3}$, Garrelt Mellema$^{4}$,\\   
         Ben Moore$^{1}$}\\  
        \vspace*{3pt} \\
 $^1$Institute for  Theoretical  Physics, University  of
           Zurich,  CH-8057, Zurich,  Switzerland\\  
       $^2$Max-Planck-Institute for Astronomy, K\"onigstuhl 17,  
           D-69117 Heidelberg, Germany\\ 
       $^3$Canadian Institute for Theoretical Astrophysics, University
                 of Toronto, 60 St. George Street, Toronto, ON M5S 3H8,
                 Canada\\  
      $^4$Stockholm Observatory, AlbaNova University Center,
      Stockholm University, SE-106 91 Stockholm, Sweden}


\date{}

\pubyear{2007}

\maketitle

\label{firstpage}


\begin{abstract}
The reionization of the Universe has profound effects on the way galaxies 
form and on their observed properties at later times. Of particular importance 
is the relative timing of the reionization history of a region and its halo 
assembly history, which can affect the nature of the first stars formed in 
that region, the properties and radial distribution of 
its stellar halo, globular cluster population and its satellite galaxies. 
We distinguish two basic cases for the reionization of a halo - internal 
reionization, whereby the stars forming in situ reionize their host galaxy, 
and external reionization, whereby the progenitor of a galaxy is reionized 
by external radiation before its own stars are able to form in sufficient 
numbers. We use a set of large-scale radiative transfer and structure 
formation simulations, based on cosmologies derived from both WMAP
1-year and WMAP 3-year data, to evaluate the mean reionization redshifts and
the probability of internal/external reionization for Local Group-like
systems, galaxies in the field and central cD galaxies in clusters. 
We find that these probabilities are strongly dependent on the underlying
cosmology and the efficiency of photon production, but
also on the halo mass. There is a rapid transition between 
predominantly external and predominantly internal reionization at a
mass scale of $\sim 10^{12}M_\odot$ (corresponding roughly to L*
galaxies), with haloes
less massive than this being reionized preferentially from distant sources. 
 We provide a fit for the reionization redshift as a
function of halo mass, which could be helpful to parameterize
reionization in semi-analytical models of galaxy formation on
cosmological scales. We find no statistical correlation between the
reionization history of field galaxies and their environment.

\end{abstract}




\section {Introduction}
\label{intro}
The lack of Gunn-Peterson troughs in the spectra of distant galaxies and QSO's 
show that the Universe has been almost completely ionized ever since redshift 
$z\sim6$. This reionization was a consequence of the ionizing radiation emitted
by the first galaxies, which started forming at $z>20$ through the
gravitational growth of the primordial density fluctuations. The recent 3-year
results from the WMAP satellite \citep{2006astro.ph..3449S} indicated that
reionization was well-advanced by $z\sim10$ and was thus fairly extended in
time. Reionization had some profound implications on star and galaxy
formation. First, stars forming in a reionized medium of low metallicity are
probably considerably less massive than stars forming in a neutral medium of
low metallicity \citep[e.g.][]{2006MNRAS.366..247J} and follow a different
evolutionary path. Second, the formation of low-mass galaxies is suppressed in
the presence of an ionizing background. The shallow gravitational wells of the
small haloes cannot hold onto their gas if the halo virial temperature is
below $T_{\rm vir}\sim10^4$~K  
\citep*[\citealt{1986ApJ...303...39D}, \citealt{2004MNRAS.348..753S}, ][]{2005MNRAS...361..405I}, while for larger haloes 
($10^4\rm K\lesssim T_{\rm vir}\lesssim10^5$~K) the gas accretion rate decreases, 
and the cooling time of the gas increases, thereby suppressing further star 
formation \citep{1992MNRAS.256P..43E,1996ApJ...465..608T,1997ApJ...478...13N,
2000ApJ...539..517B,2004ApJ...601..666D}. Therefore, reionization is a promising
candidate for reconciling the discrepancy between the number of observed low 
luminosity satellite galaxies around the Local Group and the CDM simulation 
predictions, often referred to as the 'missing satellite problem' 
\citep{1999ApJ...522...82K,1999ApJ...524L..19M, 2002ApJ...569..573T}. Third, reionization
may have truncated the formation of the blue, old globular clusters
that are still observed today \citep{2003egcs.conf..348S, 2006MNRAS.368..563M}.
Since massive galaxies will be reionized considerably before reionization is
completed in the entire IGM, it is interesting to ask when and how a galaxy of
a given mass, located in a given environment, will be reionized.

Generically, there are two ways in which a galaxy can be reionized. The
process could be driven by the radiation emanating from its own stars forming in
situ (`internal reionization') or the galaxy or its progenitors could be
overrun by the global ionization fronts created by objects which collapsed
earlier (`external reionization'). The second scenario is expected to occur
e.g. for low mass satellites that have been close to large galaxies like the
Milky Way since before reionization, while the first scenario should be
typical for very massive galaxies in the central regions of clusters. It
is less clear what scenario should be expected for intermediate-mass
galaxies. For example, a field galaxy which resides close to a massive 
cluster could have been reionized by the radiation coming from the
proto-cluster before the galaxy itself formed. The concept of a
non-uniform, density-dependent reionization redshift 
has been introduced by \citet{2004ApJ...609..474B}. They point out that
external reionization may become important in
underdense regions and voids, thus causing the relatively low
number of dwarf galaxies in these regions. 

In this work we study the reionization history of haloes and its dependence on
the halo mass and environment. To this end, we explore data from series of
large-scale structure formation and radiative transfer simulations of cosmic
reionization. We re-simulate the same initial conditions at lower resolution
which allows for tagging of the particles and thus for following the detailed
merger and accretion history of haloes from the reionization epoch to the
present. This allows us to identify the progenitors of massive cluster
galaxies, field galaxies and Local Group-like systems and simultaneously
follow both their reionization history and their mass growth history. 

In this paper, we will use two different flat $\Lambda$CDM cosmologies, the first with 
parameters ($\Omega_m,\Omega_\Lambda,\Omega_b,h,\sigma_8,n)=(0.27,0.73,0.044,
0.7,0.9,1)$ \citep[][hereafter WMAP1]{2003ApJS..148..175S} and 
($\Omega_m,\Omega_\Lambda,\Omega_b,h,\sigma_8,n)=(0.24,0.76,0.042,0.73,0.74,
0.95)$ \citep[][hereafter WMAP3]{2006astro.ph..3449S}. Here $\Omega_m$, $\Omega_\Lambda$, 
and $\Omega_b$ are the total matter, vacuum, and baryonic densities in units 
of the critical density, $h$ is the Hubble constant in units of 
100~$\rm km\,s^{-1}Mpc^{-1}$, $\sigma_8$ is the standard deviation 
of linear density fluctuations at present on the scale of 
$8 h^{-1}{\rm Mpc}$, and $n$ is the index of the primordial power 
spectrum. 

This paper is organized as follows: In \S~\ref{sec:sim} we summarize our 
simulations. In \S~\ref{sec:methods} we present our techniques used to 
select field haloes and follow the reionization histories of individual 
haloes. In \S~\ref{sec:results} we present our results for the reionization
histories of field galaxies, cD galaxies in the centers of galaxy
clusters and Local Group-like objects. In \S~\ref{sec:caveats} and
\S~\ref{sec:impli} we discuss potential caveats and discuss the
implications of external vs. internal reionization on the observable
galaxy characteristics.
Finally, in \S~\ref{sec:resdis}, we give a summary of our results.

\section{Simulations}
\label{sec:sim}

\begin{figure}
\begin{tabular}{c}
\psfig{file=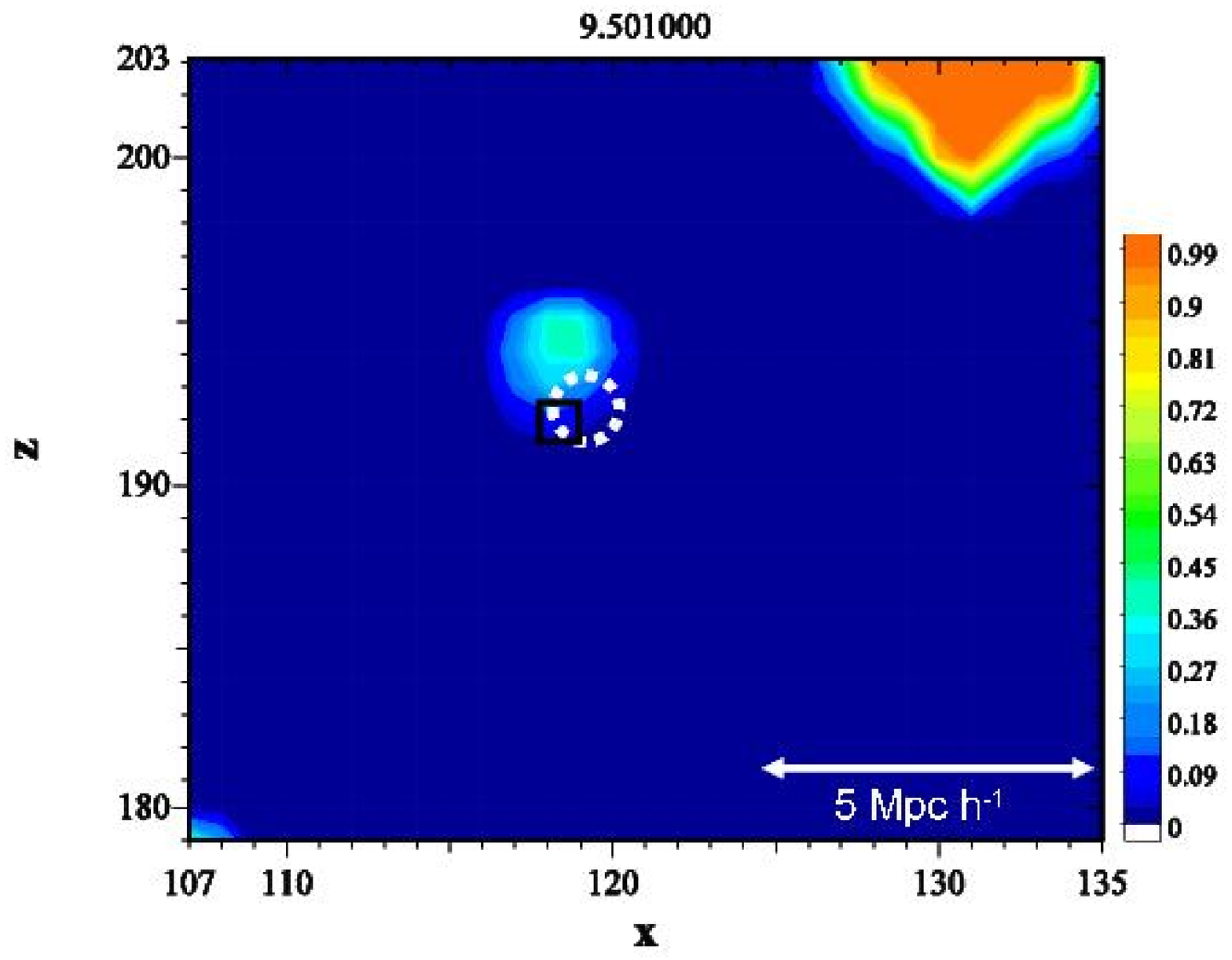,width=220pt} \\
\psfig{file=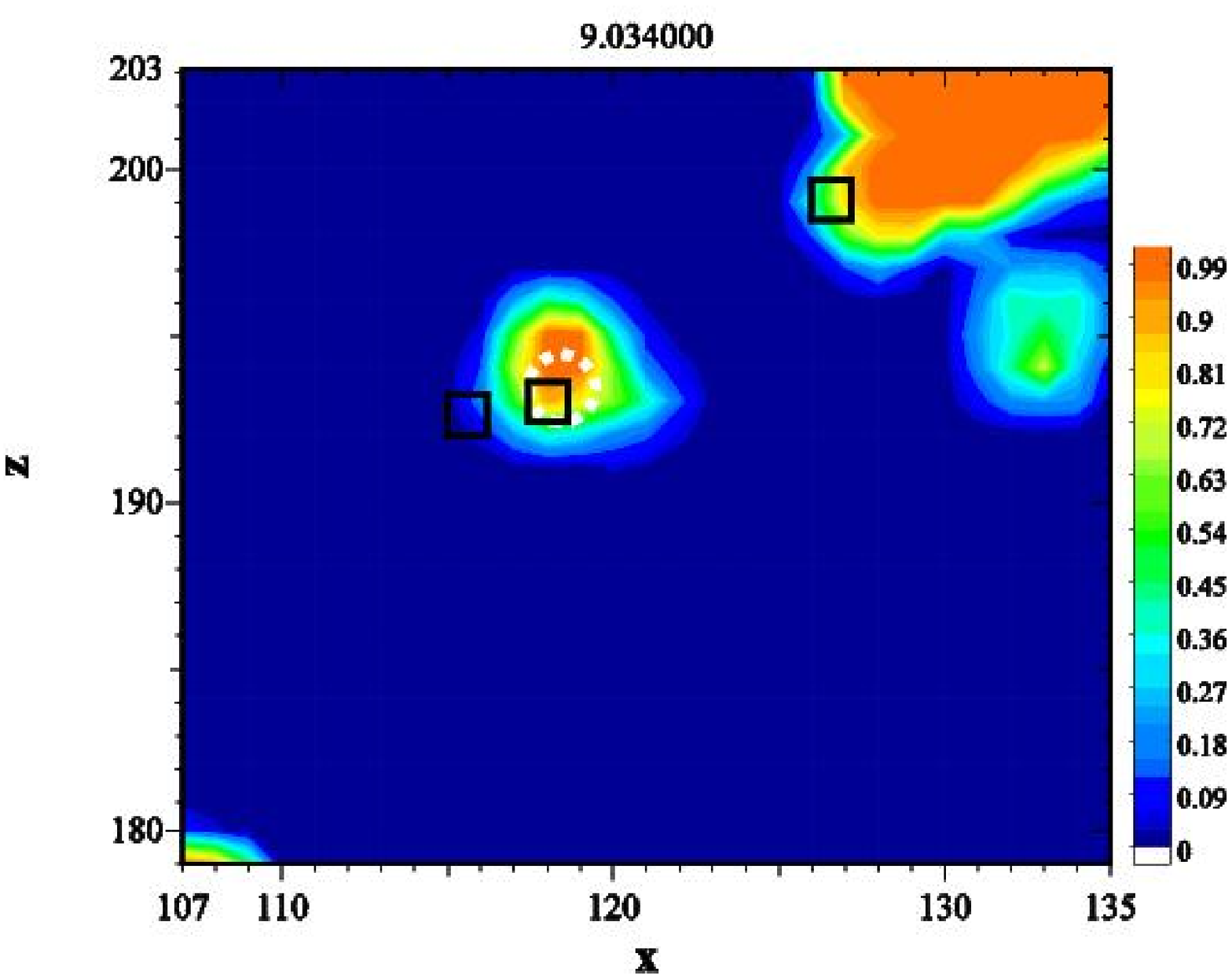,width=220pt} \\
\psfig{file=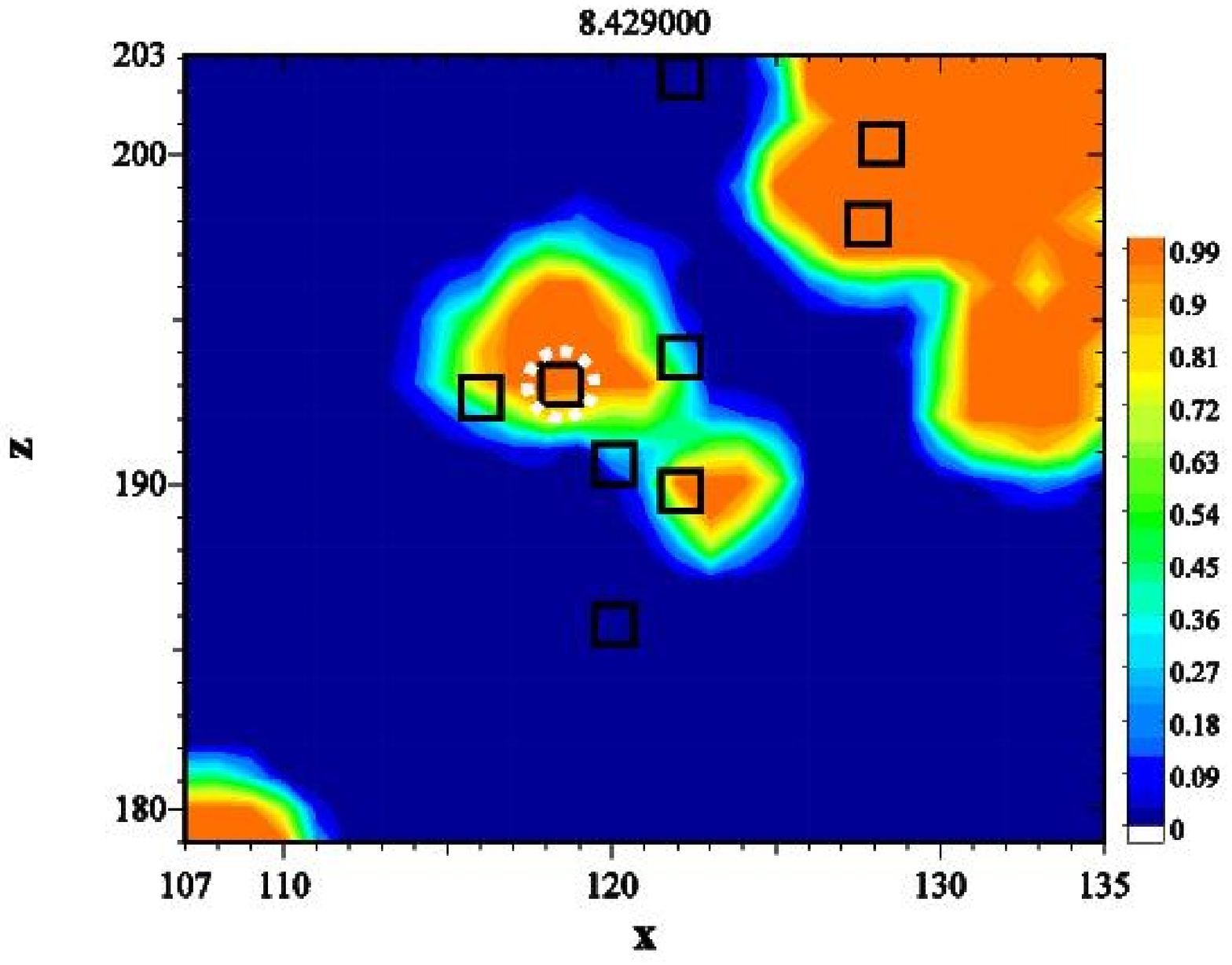,width=220pt} \\

\end{tabular}
\caption{Reionization colour plot, showing an example of internal
 reionization. Redshifts are denoted on the top of the plots. The
 reionization status of the gas is shown in colours, dark blue showing
 completely neutral gas, orange completely reionized gas.
 Black rectangles denote the position of identified haloes, the white 
 dashed circle denotes the position of
 the LG progenitor, identified by its densest particles. 
 The ionization state of the gas was identified in a slice with a
 thickness of one cell of the
 simulation outputs at different redshifts, with the position of the
 slice centered around the Local Group progenitor in
 consideration. Identified haloes are shown if they are within a box
 of volume $20^3$ cells centered around the Local Group progenitor
 (i.e. can be up to 10 cells offset from the slice shown in the
 pictures).}
\label{fig:maps2}
\end{figure}
\begin{figure*}
\begin{tabular}{cc}
\psfig{file=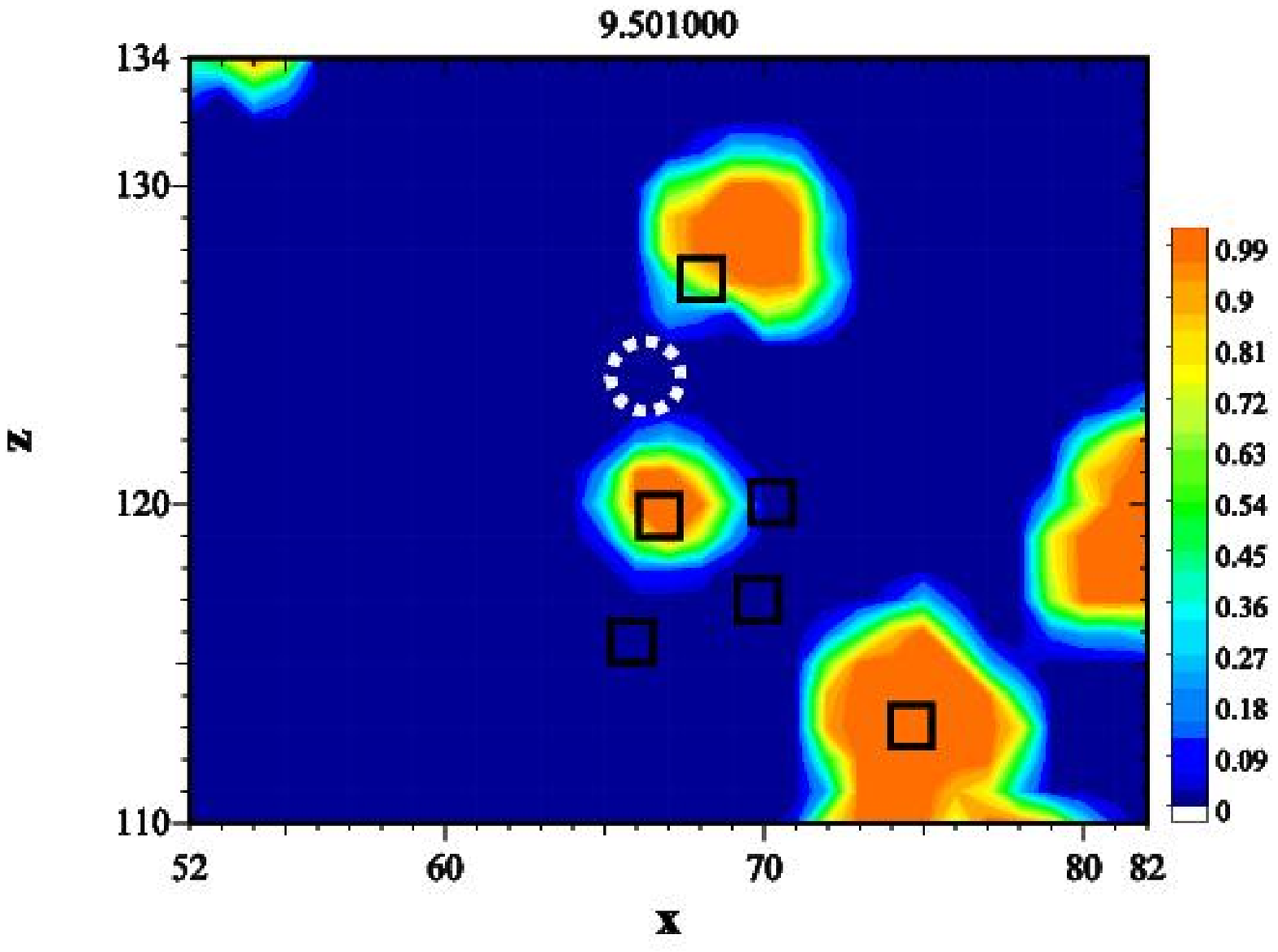,width=220pt} &
\psfig{file=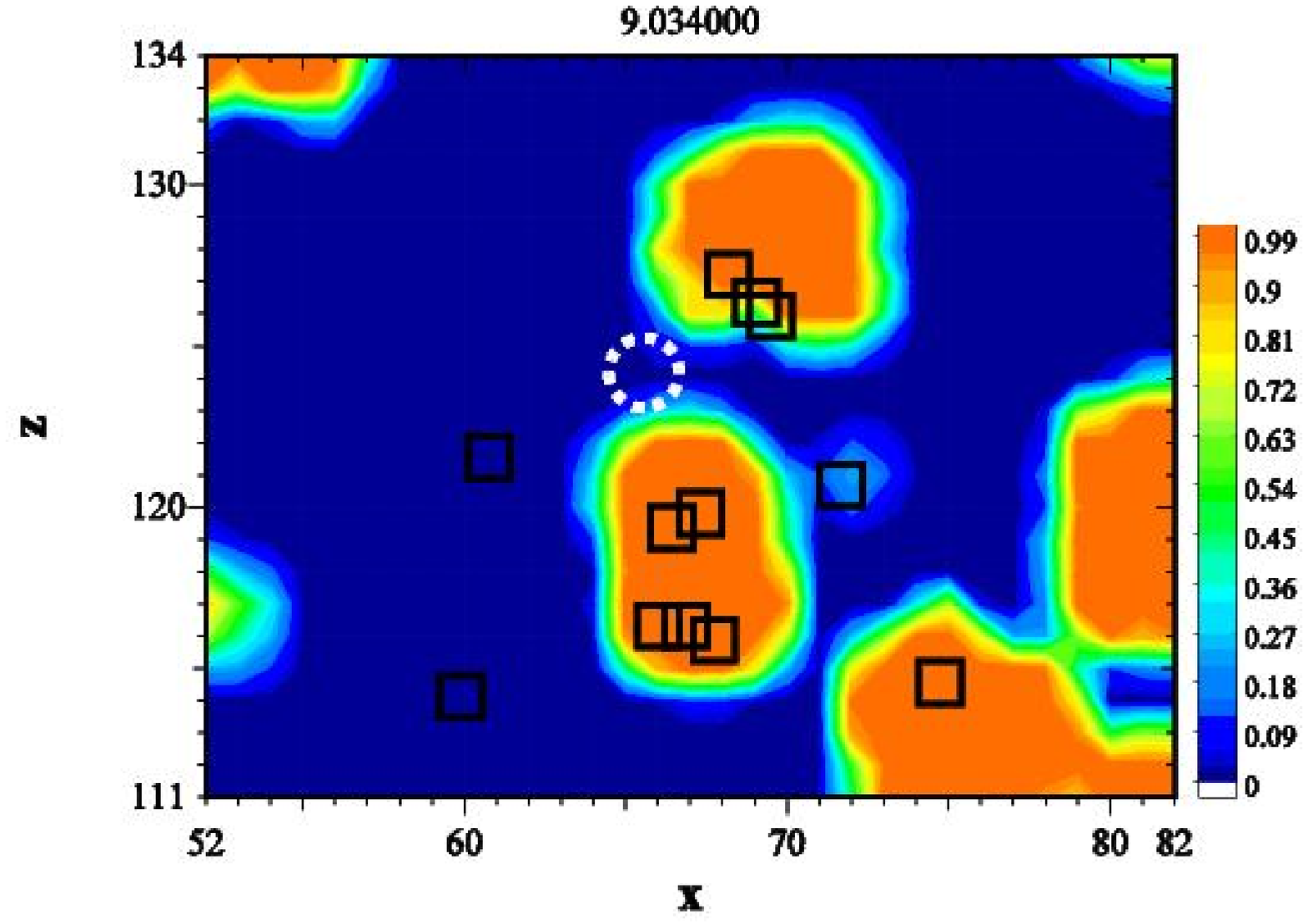,width=220pt} \\
\psfig{file=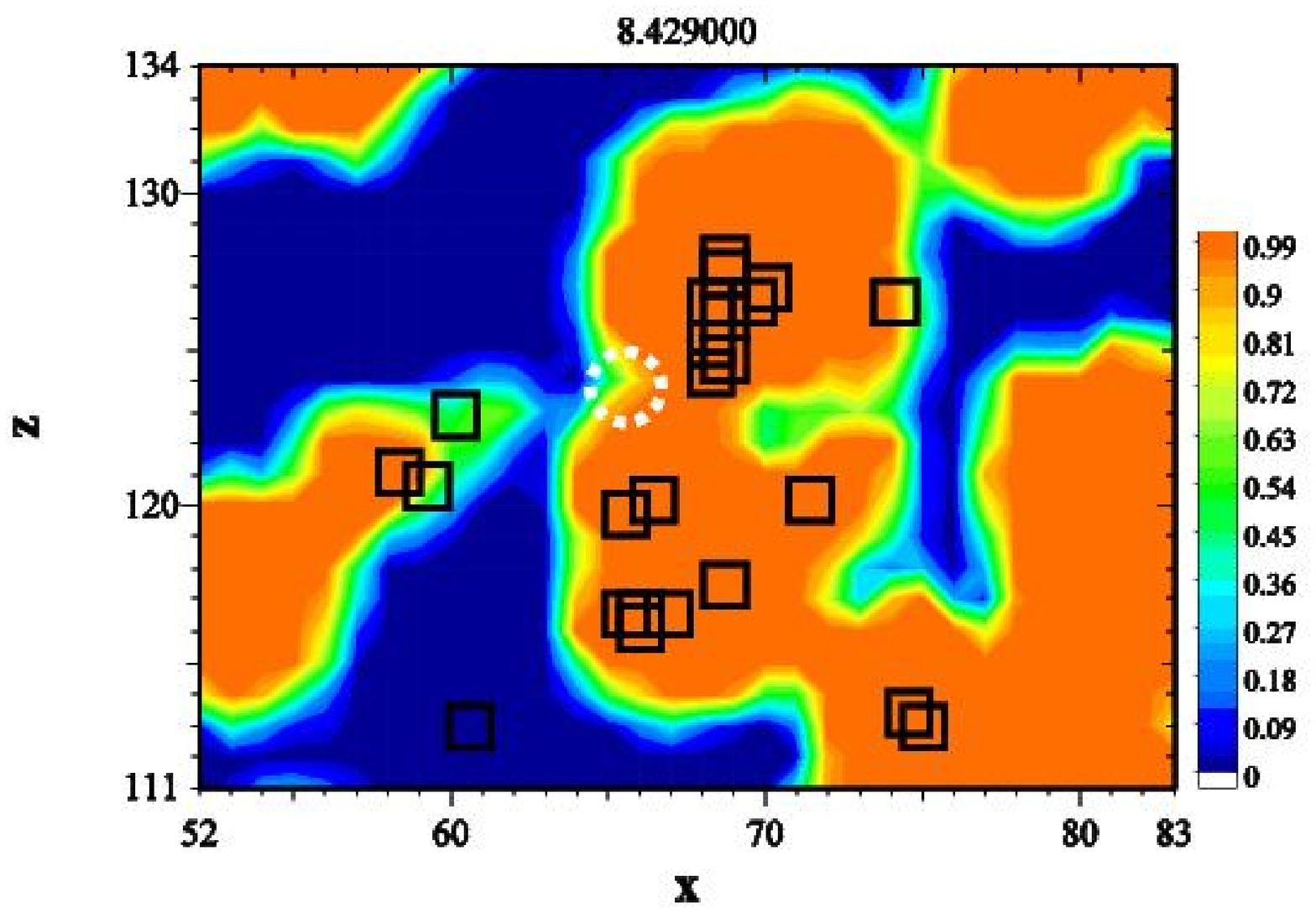,width=220pt} &
\psfig{file=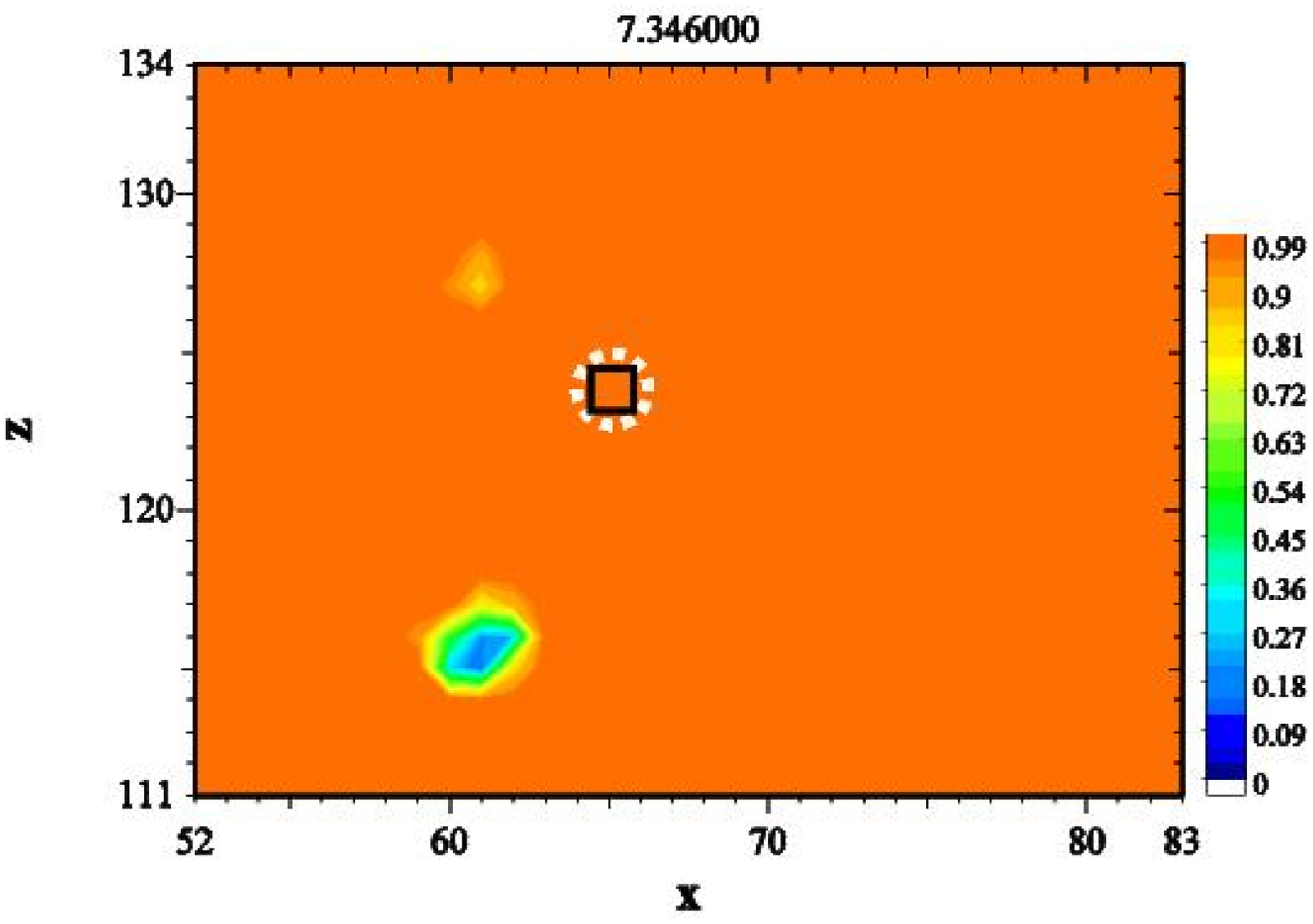,width=220pt} \\

\end{tabular}
\caption{ Reionization colour plot, showing an example of external
 reionization. Symbols/Colours have the same meaning as in
  Figure~\ref{fig:maps2}. 
At $z$=7.346, the number of haloes is so large that we do
 not show them anymore in the plot.}
\label{fig:maps}
\end{figure*}

\begin{table}

\caption{Overview of the three different codes and the resolutions
  used in our WMAP3 simulation (WMAP1 is similar). Note that the
  spatial resolution in PMFAST is better by at least an order of
  magnitude as far as the position of halo centers are concerned. In
  the PKDGRAV simulation, the spatial resolution is not clearly
  defined and was assumed to be equal to the force resolution.}
\begin{center}
\begin{tabular}{|llll|}
\hline
& code & mass res. &  spatial res.  \\

\hline\hline
high res & PMFAST & 2 $\times 10^7$ $\Msunh$ & $\sim$ 30 $h^{-1}$kpc\\
low res & PKDGRAV & 9.96 $\times 10^8$ $\Msunh$ & $\sim$ 10 $h^{-1}$kpc\\
RT & C$^2$-ray & -  & $\sim$ 0.5 $h^{-1}$Mpc\\
\hline
\end{tabular}
\end{center}
\label{code_params}
\end{table}

Our simulations follow the evolution of a comoving simulation volume
of $(100\,h^{-1}\rm Mpc)^3$. Our basic methodology and simulation 
parameters are described in detail in \citet{2006MNRAS.369.1625I,
21cmreionpaper,selfregulated}. Here we provide just a brief summary, in
addition to a detailed discussion of the elements which are first
presented in this work. See Table~\ref{code_params} for listing of the 
codes used in this work and their effective mass and spatial resolutions.

\subsection{Structure Formation Simulations}
\label{DMonly}

We start by performing a very large dark matter only simulation of early 
structure formation, with $1624^3\approx4.3$ billion particles and $3248^3$ 
grid cells\footnote{$3248=N_{nodes}\times(512-2\times24)$, where
  $N_{nodes}=7$ (with 4 processors each), 512 cells is the Fourier transform
  size and 24 cells is the buffer zone needed for correct force matching on
  each side of the cube.} 
using the code PMFAST \citep{2005NewA...10..393M}. This allows 
reliable identification (with 100 particles or more per halo) of all haloes 
with masses $\sim2 \times 10^9 \Msunh $ or larger. We compile and save the halo
catalogues, in up to 100 time slices starting from high redshift
($z\sim30$) to the observed end of reionization at $z\sim6$. 
In what follows, we will refer to this simulation as HR
(high-resolution) simulation. 
This HR-simulation and in particular the halo catalogues based on it
are used as the framework for our radiative transfer (RT) simulation
which will be explained in detail in section \ref{RT}.
In order to maximize the mass resolution, 
the dark matter particles are not labelled in the HR simulation,
 which does not allow us to identify the progenitors of 
any given structure at $z=0$. We resolved this problem by re-simulating the same 
initial conditions at lower resolution (called LR simulation
hereafter) which allows for tagging of the particles.
 This enables us to identify the progenitors of 
haloes at $z$=0 and to
simultaneously follow their reionization and mass growth histories. 
To obtain the LR simulation, the original HR PMFAST initial conditions 
(with $1624^3$ particles) were generated again using the same random seeds. They 
were then coarsened to $406^3$ particles by averaging the properties (positions 
and velocity components) of every $4^3$ neighboring particles. We evolved these 
coarsened initial conditions from $z=100$ to $z=0$ using PKDGRAV, a fast parallel 
tree code written by Joachim Stadel and Tom Quinn
\citep{2001PhDT........21S}.

\subsection{The Radiative Transfer Simulations}
\label{RT}
We proceed by applying our radiative transfer code to the
HR-simulation described above. Since radiative transfer simulations at the
full grid size of our N-body simulations are still impractical, we follow the
radiative transfer on coarser grids, of sizes $203^3$. 
All identified haloes are assumed to be sources of ionizing radiation and 
each is assigned a photon emissivity proportional to its total mass, $M$, 
according to
\begin{equation}
  \dot{N}_\gamma=f_\gamma\frac{M\Omega_b}{\mu m_p t_s\Omega_0},
\end{equation}
where $t_s$ is the source lifetime, $m_p$ is the proton mass, $\mu$ is the 
mean molecular weight and $f_\gamma$ is a photon production efficiency which 
includes the number of photons produced per stellar atom, the star formation 
efficiency (i.e. what fraction of the baryons are converted into stars) and 
the escape fraction (i.e. how many of the produced ionizing photons escape the 
haloes and are available to ionize the IGM).
Since we do include in the halo catalogue only haloes defined by more than 100
particles, this criterion introduces a minimum mass for haloes in order to act
as ionizing sources. For our choice of the simulation parameters this
corresponds to a mass threshold of $M_{th} = 2.5 \times10^9\Msunh$ (WMAP1) and
$M_{th} = 2.2\times10^9\Msunh$ (WMAP3). Note that this mass corresponds
roughly to the halo mass above which a galaxy forming in it is largely 
unaffected by reionization \citep[see][for discussion and
references]{selfregulated}. 

The radiative transfer is followed using our fast and accurate ray-tracing
photoionization and non-equilibrium chemistry code C$^2$-Ray
\citep{methodpaper}. The code has been tested in detail for correctness and
accuracy against available analytical solutions and a number of other
cosmological radiative transfer codes \citep{methodpaper,comparison1}. The
radiation is traced from every source on the grid to every cell using
short-characteristic ray-tracing. The runs and notation are the same as the
ones presented in \citet{2006MNRAS.369.1625I,21cmreionpaper,selfregulated}.
These include simulations with both WMAP1 and WMAP3 background cosmologies. 
The runs for a given cosmology share the same underlying N-body simulation 
(but with different random realizations for WMAP1 and WMAP3), and each adopts
different assumptions about the source efficiencies and the sub-grid level of
gas density fluctuations. In part of the simulations, a sub-grid gas clumping
is included, $C(z)=\langle n^2\rangle/\langle n\rangle^2$, which evolves with
redshift according to
\begin{equation}
C_{\rm subgrid}(z)=27.466 e^{-0.114z+0.001328\,z^2}.
\label{clumpfact_fit}
\end{equation}
in WMAP1 cosmology and as
\begin{equation}
C_{\rm sub-grid}(z)= 26.2917e^{-0.1822z+0.003505\,z^2}.
\label{clumpfact_fit3}
\end{equation}
for WMAP3 cosmology. These fits were obtained from another set of two
high-resolution PMFAST N-body simulation, with box sizes $(3.5\,\rm
h^{-1}~Mpc)^3$ and a computational  
mesh and number of particles of $3248^3$ and $1624^3$, respectively. 
The inclusion of gas clumping results in a slower propagation of the
reionization fronts and a delay of the final overlap.

In what follows, we will concentrate on three cases, two 'extreme' models and
an intermediate one. The extreme cases (in terms of $z_{\rm ov}$, the
redshift at which the final overlap is reached), are the following:
WMAP1-f2000 (WMAP1, $f_{\gamma} =2000$, no sub-grid gas clumping) and
WMAP3-f250C (WMAP3, $f_{\gamma} =250$, with sub-grid gas clumping). These two
simulations should  
roughly bracket the allowed range in cosmological and reionization parameters. 
Among all of our radiative transfer simulations, the WMAP1-f2000 simulation
has the earliest redshift of overlap, $z_{\rm ov}$=11.3, while simulation
WMAP3-f250C has the latest overlap, $z_{\rm ov} \sim$ 6.5. We add a
simulation in between those two cases: WMAP1-f250C (WMAP1, $f_{\gamma}
=250$, with sub-grid gas clumping), resulting in $z_{\rm
  ov}$=8.2. This allows us to study
the effect of changing only the cosmology or changing only the
reionizing photon production efficiency. 
The lower efficiency of
$f_{\gamma} =250$ appears more realistic at present, and the corresponding
cases match the late end of reionization at $z\sim7$ well. We note that the 
delayed structure formation due to the lower power spectrum normalization,
tilt and lower matter density in the favoured WMAP3 cosmology naturally
predicts the lower integrated Thomson scattering optical depth the same
measurements found, and thus do not seem to require a lower reionizing photon 
production efficiency \citep[][but see also
\citealt{2006ApJ...650....7H, 2007astro.ph..2323S}]{2006ApJ...644L.101A}.  

For the WMAP1 simulations, transmissive boundary conditions on the ray-tracing
were used. As a consequence, haloes close to the boundaries of our
computational volume will be somewhat less affected by external radiation,
particularly towards the end of reionization. Thus, in order to be
conservative, we do not include in our analysis any haloes that are within 10
$h^{-1}\rm Mpc$ from the simulation boundary. The WMAP3 simulation utilized
periodic boundary conditions, which allows for all haloes to be used in our
analysis.

\section{Methods}
\label{sec:methods}
\subsection{Matching the HR and LR Simulations}

In order to investigate the reionization and formation history of a given
halo, we first identify it at $z$=0 in our LR simulation, using a
Spherical Overdensity Halo Finder (see \citealt{2006astro.ph..8157M}
for a detailed description). We then track the particles of
the halo in consideration back to higher redshift in the LR simulation
and match them to the HR simulation in order to analyze its formation and
reionization history.  
This can only be done if the density distribution in the HR and LR
simulation agree well; since two quite different methods and resolutions are
used for them, this merits checking and evaluating the closeness of the match
between the two. If they match well, the location of the haloes
which can be identified in the LR simulation should be similar to the
location of the most massive haloes in the HR simulation (where the
total number of identified haloes at any redshift is much higher than
in the LR simulation).
As a test, we identify for each halo in the LR simulation which is
found by our Spherical Overdensity Halo Finder the nearest halo in
the HR simulation which has a similar mass (within a factor of 2).
At $z$=6 in the WMAP3 simulation, we find that the median distance 
between the haloes in the two catalogues is 0.3 Mpc $h^{-1}$, with a 
similar result for the WMAP1 simulation. Since this distance is below 
the grid size of the RT code, we consider this reasonable agreement for 
our current purposes. However, for around 3 percent of all cases, no 
nearby halo with similar mass is identified. This is not surprising 
given the fact that we use two different halo finders and have a large 
difference in resolution. 
Note that the median distance and the number of unmatched haloes
decreases with increasing redshift. This reflects the fact that the DM
distribution in the HR and LR simulation becomes more similar at high $z$,
since the effects of different force resolution, etc., are diminished.
As all of our comparisons between LR and HR simulations are done at
times when the Universe is not yet completely reionized, i.e. $z >$ 6,
we consider
the median distance and the fraction of unmatched haloes quoted above
an upper limit. 

In what follows, we will explain in more detail how the formation and
the reionization history of a given $z=0$ halo is investigated.
We mark the particles 
that will end up in a given halo at $z=0$ and track them back to 
simulation outputs at higher redshifts. At any redshift, we identify the 
center of the halo progenitor as the densest particle (smoothing over the 
32 closest neighbours) in the marked set and measure the fraction of reionized 
gas in the central cell and the 26 neighbouring
cells, using the radiative transfer grid consisting of $203^3$
cells. In this way, an offset of one radiative transfer (RT) cell (corresponding to
$\sim0.5h^{-1}$Mpc) in each dimension is allowed between the center
of reionization in the HR-simulation, and the location of the densest
particle in the LR-simulation. If the ionized fraction in any of 
those cells is higher than 
a threshold value of $x_{\rm th}=0.7$, we define the halo as reionized and the 
corresponding reionization redshift $z_{r}$ is set equal to the current output 
redshift. The particular value of $x_{\rm th}$ has a negligible impact on the 
final results thanks to the sharp transition across the ionization
fronts (see Fig. \ref{fig:maps2} and \ref{fig:maps}).

Next we check  at which redshift the most  massive progenitor of the
halo in consideration first reaches a halo mass of $M_{th}$ (as defined in
\S~\ref{DMonly}), and define  this as redshift of formation,  $z_{f}$. To 
do this, we again trace back the halo progenitor particles and identify 
the densest particle at each redshift. Again allowing for an offset of
one cell, we consider a halo as formed if it is identified by our halo
finder based on the HR simulation. Only for less than 2 percent of all
haloes, no formation redshift  could be identified.  They were removed
from the sample.

Ionizing radiation in the ray tracing simulation can, by definition,
only be produced by haloes that have been identified; this allows us to
clearly separate internally reionized haloes from externally reionized
ones. If $z_r > z_f$ ($z_r \le z_f$), the halo has been reionized externally
(internally). For illustrative purposes, in Figures \ref{fig:maps2} and
\ref{fig:maps} we show the evolution of the ionization fronts around an
internally and an externally reionized Local Group-like object, respectively.

Our method can occasionally fail if the halo progenitor under consideration
has a very large extent (i.e. is larger than two cells) and has two dense
centers. If one of those is identified as densest particle by our smoothing
algorithm, while the other is recognized by the high-resolution halo 
finder, external reionization will be always assumed. Eye inspection 
indicates that this happens in less than roughly $ 10 \%$ percent of all cases
we have identified as external reionization. On the other hand, if the
reionization and the formation of a halo occur in-between two timesteps, we
always assume internal reionization. Although external reionization is also
possible it is unlikely. We did not find a way to 
solve these two issues, but we expect that their impact on our results
should be modest, since these effects will both only occur for a small
fraction of our haloes.

\subsection{Sample Selection}
We have identified dark matter haloes at $z$=0 in our LR simulation
using our Spherical Overdensity Halo Finder. From this halo catalogue we
selected three sets of objects for further consideration: 1) field haloes, 
2) central objects of massive clusters, likely to correspond to 
central cD galaxies and 3) Local Group-like systems. As a subset of
the field haloes, we also selected a sample of haloes likely to host
an L* galaxy.

The sample sets of such systems were chosen as follows. The field halo samples 
consist of all the haloes 
with masses between $1.8\times 10^{11} \Msunh$ and  $3.7\times10^{13}
\Msunh$ that have been found by the Spherical Overdensity Halo Finder
in both simulations. Such haloes are always isolated in the sense that
they are not subhaloes in a larger group or cluster. 
In the  WMAP1-f2000 and the WMAP1-f250C simulation, haloes with
distances closer than 10 $h^{-1}$ Mpc from the edges of the simulation
box were removed from the sample. In
this way, we have identified 9683 haloes in the WMAP3-f250C
simulation, and 7346 haloes in the WMAP1-f2000/WMAP1-f250C simulations.
A sample of haloes likely to host an L* galaxy was chosen as follows. 
According to \citet{2002MNRAS.333..133M},
  L* - galaxies have a $b_J$-luminosity of $1.1 \times 10^{10}$
  $h^{-1}\Lsun$. A central  or isolated L*
  galaxy is therefore most likely to be found in a halo with
  log($M$)=12.2 [$\Msunh$] in a  WMAP3, and in a halo with mass
  log($M$)=12.3 [$\Msunh$] in a
  WMAP1 cosmology, following the relation between $L_{cen}$ and halo mass in
  \citet{2007MNRAS.tmp..139V}. The difference in the two cosmologies is
  mainly due to the different halo clustering properties. 
 We select for our L* galaxy sample haloes with mass
 log($M$)=12.1-12.3 [$\Msunh$] for the simulation based on WMAP3, and with
 log($M$)=12.2-12.4 [$\Msunh$] for the simulations based on WMAP1. We find 437 such
 haloes in the two simulations based on the WMAP1 cosmology, and 575
 haloes in the WMAP3-f250C simulation.

The second set of haloes consists of the central objects in the 20 most
massive galaxy clusters in each simulation. These massive  
objects are likely to correspond to central cluster cD galaxies. 
Finally, the third halo sample consists of all Local Group-like systems in our
volume. These are identified as binary systems of dark matter haloes
satisfying the following criteria \citep*[see][for
details]{2005MNRAS.359..941M}: 
\begin{enumerate}
\item The mass of each of the two haloes is between $4 \times 10^{11}$
  and 
$5 \times 10^{12} \Msunh$.
\item The distance between the two haloes is below 700 $ h^{-1}$ kpc.
\item The two haloes are moving towards each other.
\item There is no massive neighbour with a mass above $5 \times
  10^{11}$ closer than 2.1 $h^{-1}$ Mpc.
\end{enumerate}


\section{Results}
\label{sec:results}
\subsection{Reionization Histories of Galaxies of Different Mass and
  Environment}

\begin{table}

\caption{Summary of the results. Listed are the mean redshift of formation,
  $z_{f}$, the mean starting redshift of the halo reionization, $z_{r}$, and
  the mean redshift at which 70 percent of the final halo mass is reionized,
  $z_{70\%}$, the fraction of objects reionized externally, $f_{\rm ext}$,
  the average time interval between formation and reionization, $\Delta
  t\equiv t_f-t_r$ (in Myr), and $\Delta
  t*$, which is the same time difference in units of the
  Hubble time when 50 $\%$ of the mass in the universe is ionized, for all of our
  simulations, as labelled. If $\Delta t$ is negative, haloes are on
  average reionized {\it before} formation. } 
\begin{center}
\begin{tabular}{lllllll}
\hline
& $z_f$ & $z_r$ & $z_{70\%}$ & $f_{\rm ext}$ &  $\Delta t$&$\Delta t^*$\\
\hline\hline

{\bf WMAP1-f2000}\\
\hline
field haloes & 10.8 & 13.4 & 13.1 & 0.79 & -114   & -0.237\\
L* halo sample & 12.6 & 13.6 & 12.85 & 0.62 & -36 &-0.075 \\
central cDs & 16.7 & 15.5 & 14.1 & 0.0 & 26       & 0.055\\
LG sample & 12.8 & 13.3 & 12.9 & 0.44 & -21   &-0.044       \\
\hline 
{\bf WMAP1-f250C}\\ 
\hline 

field haloes & 11.0 & 11.7 & 10.4 & 0.4 & -33  & -0.053\\
L* halo sample & 12.6 & 12.4 & 10.9 & 0.17 & 8 & 0.013\\
central cDs & 16.7 & 15.5 & 11.8 & 0.0 & 26 & 0.042\\
LG sample & 12.8 & 12.4 & 10.8 & 0.1 & 13 & 0.021\\

\hline
{\bf WMAP3-f250C}\\ 
\hline
field haloes & 8.2 & 8.9 & 8.2 & 0.56 & -60  & -0.062\\
L* halo sample & 9.3 & 9.4 & 8.6 & 0.27 & -9 & -0.009\\
central cDs & 13.0 & 12.6 & 9.6 & 0.0 & 15 &  0.016\\
LG sample & 9.1 & 9.2 & 8.3 & 0.24 & -8      & -0.008\\
\hline
\end{tabular} 
\end{center}
\label{wmap1table}
\medskip
\end{table}

On average, reionization {\it predates} formation for typical field
haloes in all three simulation cases that we have considered.
We compare the redshift of formation, $z_{f}$, the starting redshift of
the halo reionization, $z_{r}$, and the redshift at which 70 percent of 
the final halo mass is reionized, $z_{70\%}$, for all our halo samples
(field haloes, haloes likely to host an L* galaxy, haloes likely to
host a cD galaxy, LG like haloes) in Table~\ref{wmap1table}. We note that by
definition $z_r > z_{70\%}$, and especially for massive haloes these redshifts
could differ significantly  
since the material that will end up in the final halo is generally spread 
out over a very large region in space, and therefore a fairly long time could 
pass from start to finish of the reionization of that halo's material.
As mentioned before, $z_{f}$ is the redshift at which the largest
progenitor of the halo reaches $M_{th}$.

\begin{figure}
\centerline{\psfig{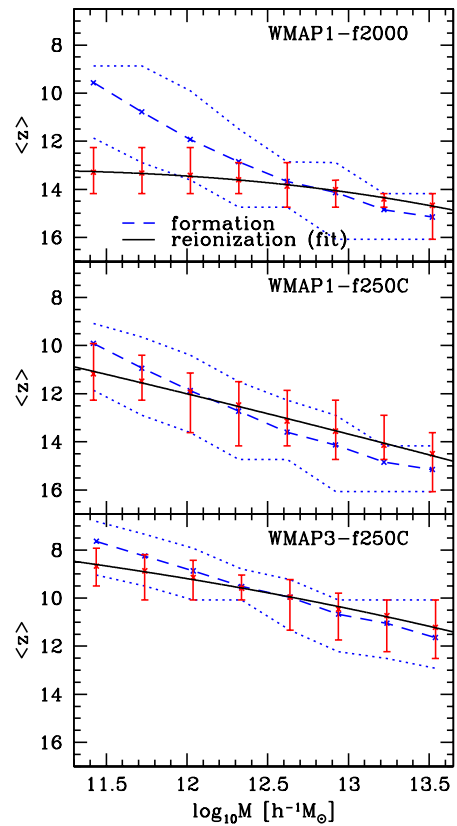}}
\caption{The average formation and reionization redshift (calculated converting redshift to
  lookback time and back) as a function of the halo mass for the WMAP1-f2000
 (top panel), the WMAP1-f250C(middle panel) and the WMAP3-f250C (bottom panel) simulation. Errorbars/dotted
  lines denote the 68 percent confidence level. The black solid line is a
  fit to the average redshift of reionization, as explained in the
  text. Note that formation redshift here means the redshift at which
  M $>$ $M_{th}$. }
\label{fig:zav}
\end{figure}

We find that external reionization occurs for roughly 80\% of all
field haloes in the WMAP1-f2000 simulation, for 40 \% in the WMAP1-f250C
simulation and for $\sim 55 \%$ in the WMAP3-f250C simulation. Apparently,
our ``intermediate case'' with a redshift of overlap in between the
two other cases, has the lowest number of externally reionized haloes.
This is caused by an interplay of two different effects. First, the
higher the reionizing photon production, the more effective is a halo
in reionizing its surroundings. If $f_{\gamma}$=2000, each halo 
quickly reionizes a large volume, containing a mass $\sim2000$ times the halo mass, 
as recombinations are not very important. This explains the very
high fraction of external reionization in the WMAP1-f2000 case, which
indicates that in this case, few early collapsing haloes are responsible for most of
the reionization of the matter in the universe.
The influence of the cosmological framework is more complicated. As
pointed out by \citet{2006ApJ...644L.101A}, structure formation is
delayed by a factor of $\sim$ 1.4 in
(1+$z$) in a WMAP3 cosmology compared to a
WMAP1 cosmology. This means that structure formation becomes more
extended in time in a WMAP3 cosmology, i.e. the time of formation of
low and high mass haloes are further apart. In order for external
reionization to occur, the radiation of an early collapsing halo has
to reach other haloes before they have started to form. This window in
time will be shorter in a WMAP1 cosmology than a WMAP3 cosmology,
which explains the lower external fraction in the former case. 
This means that for a WMAP3 cosmology and a higher reionizing photon
production efficiency, the external fraction would be even above 80 \%.

Another important outcome of our analysis is that differences in formation
time cannot be directly translated into a difference in reionization time
(cfr. Table~\ref{wmap1table}). 
Due to the contribution of external reionization, field galaxies ``catch up'' 
with central cD galaxies when it comes to reionization.
The time elapsed between formation and reionization ($\Delta t$) depends both
on the cosmological model and on the reionization parameters. 
In the early-reionization scenario WMAP1-f2000, $\Delta t$ for field
haloes and LG-like objects is always negative and maximal
in absolute value, which means that these objects form a fairly long
time after they have been reionized.  
Field haloes are reionized on average noticeably earlier in WMAP1-f2000 model 
than in the extended reionization scenarios. 
This is to be expected considering the much faster and
earlier reionization in this case, allowing for significantly fewer field haloes to
form before reionization. Some L* haloes in the WMAP1-f2000 simulation also
do not manage to reach a mass of $M_{th}$ before their reionization. Thus the coeval formation
and reionization of L* haloes observed in the other two simulation cases
is broken for early reionization.
WMAP1-f250C and WMAP3-f250C share the same reionization
parameterization. Thus, differences in 
$\Delta t$ are due to the different cosmological models. 
The characteristic time scale for structure formation is roughly the 
current Hubble time, $t_H(z)$. So, in an attempt to scale-out the cosmological dependence of 
$\Delta t$, we divided it by the Hubble time $t_{H, 50\%}$ when the ionized fraction (in mass) is 50\% 
to obtain $\Delta t^*$=$\Delta t$/$t_{H, 50\%}$. 
Comparing the extended reionization scenarios, WMAP1-f250C and WMAP3-f250C, 
we see that $\Delta t^*$ is very similar for all objects. 

Haloes likely to host massive cD galaxies in centers of clusters are always internally 
reionized in all the different simulations. This reflects the strong inside-out
nature of reionization. Reionization always starts from the highest
density peaks, inside which cD haloes reside, and thus these are reionized
early, regardless of the details of the reionization parameters and the
background cosmology.   

In summary, the timing of reionization compared to formation for L* and
field haloes is sensitive to the reionization parameters (ionizing
efficiency of the sources and gas clumping) and rather insensitive to the
background cosmology, while in contrast the reionization history of central cD 
galaxies is largely independent of either the reionization parameters, or the
cosmology.

In Fig. \ref{fig:zav}, we show the mean values of $z_{r}$ and $z_{f}$
as a function of the
halo mass at $z=0$. Averages are calculated by converting redshift to lookback
time, averaging and converting the result again back to redshift. The
reionization redshift as a function of halo mass at $z$=0 can be fitted
by a simple polynomial:
\begin{equation}
\langle z_r \rangle =ax+bx^2+cx^3
\end{equation}
where $x$=log$(M/M_{\odot})$, expressed without $h^{-1}$.
We list the relevant values for $a$, $b$, $c$ in Table~\ref{fit}.

\begin{table}

\caption{Fitting parameters for $\langle z_r \rangle$ (M), with $\langle z_r \rangle =ax+bx^2+cx^3$,
where $x$=log($M/M_{\odot}$)}
\begin{center}
\begin{tabular}{|l|lll|}
\hline
& $a$ &$ b$ & $c$ \\
\hline\hline
WMAP1-f2000 & 5.079 & -0.599 & 0.022\\
WMAP3-f250C & 0.187 & 0.079 & -0.001 \\
WMAP3-f250C & 1.592 & -0.167 & 0.008\\
\hline
\end{tabular}
\end{center}
\label{fit}
\end{table}
There are clear
trends for the redshift of formation and reionization as a function of
halo mass. 
For lower mass haloes, reionization
typically happens prior to their formation (external reionization), but for
higher mass haloes, formation tends to predate reionization (internal
reionization). As mentioned above, \citet{2006ApJ...644L.101A} found that 
structure formation is
delayed by a factor of 1.4 in (1+$z$) in a WMAP3 cosmology. We roughly
find a similar change in the average formation time of a halo,
comparing WMAP1 and WMAP3 cosmologies. The 68 percent confidence level
around average
formation redshift is
considerably larger for a WMAP1 cosmology than for a WMAP3
cosmology at low masses. However, this simply reflects the fact that a similar
interval in time corresponds to a much larger interval in redshift at higher
redshifts. Measured in lookback time, the scatter around the average
formation redshift is very similar in the two cosmologies and even
slightly higher for WMAP3 at low masses.

\begin{figure}
\centerline{\psfig{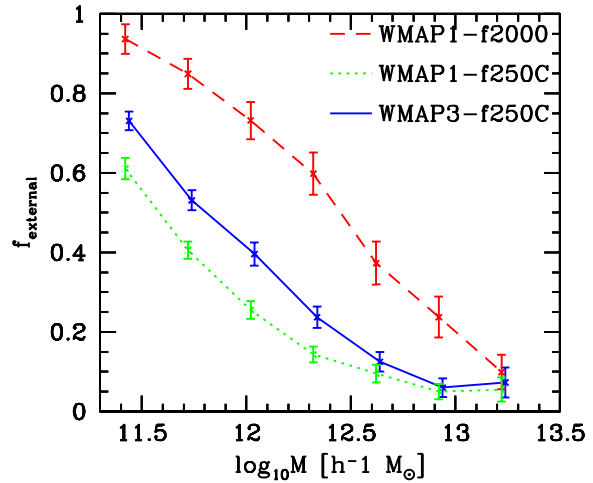}}
\caption{The fraction of externally ionized haloes as a
  function of halo mass respectively for the three simulations considered.}
\label{fig:frac1}
\end{figure}

In Fig. \ref{fig:frac1}, we show the fraction of
externally reionized haloes as a function of halo mass. 
It is apparent that the probability of external reionization is strongly dependent on
mass, mainly because the formation redshift and the mass of a halo
are correlated, and haloes which form later  have a higher
probability of being externally reionized. We note that even if on
average a halo 
forms after it vicinity has been reionized, the
fraction of externally reionized haloes can still be rather low. The
reason for this is that for many internally reionized haloes, the redshift of
formation and the redshift of reionization are the same or at least
very close. Thus, even a
relatively small fraction of externally reionized haloes can shift the mean
reionization redshift so that it predates the formation redshift.

\begin{figure}
\centerline{\psfig{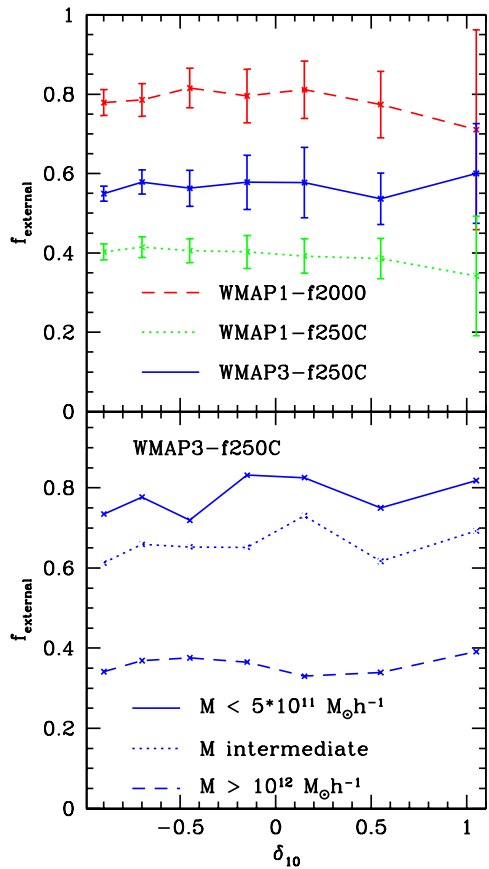}}
\caption{Top Panel: The fraction of externally ionized haloes as a
  function of the overdensity by measuring only the mass in collapsed haloes with masses above
  $5\times 10^{12}$. The overdensity is calculated with respect to the mean
  matter density in the simulation box. Results for all three simulations
  are shown. Errorbars denote the 1-$\sigma$ error.
  Bottom Panel: Same as shown top panel, for the
  WMAP3-f250C simulation only, but for three different bins regarding
  the mass of the halo in consideration. The way the surrounding
  overdensity is measured remains unchanged. For clarity, no errorbars
  are shown.
}
\label{fig:frac2}
\end{figure}

\begin{figure}
\centerline{\psfig{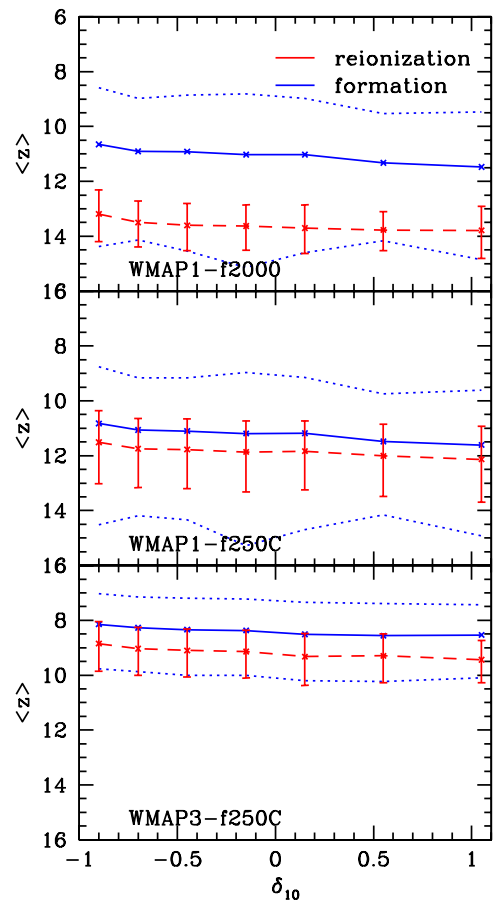}}

\caption{The average formation and reionization redshift (calculated converting redshift to
  lookback time and back) as a function of the overdensity measured
  only for the mass in collapsed haloes with masses above 5 $\times 10^{12}$, for the WMAP1-f2000
 (top panel), WMAP1-f250C and the WMAP3-f250C (bottom panel)
  simulation. Dotted lines/errorbars denote the 1-$\sigma$ error.}
\label{fig:tim2}
\end{figure}

Generally, one might expect that the probability of external reionization
of a field galaxy could also depend upon the local overdensity around it, since
the latter is closely related to the abundance of sources of ionizing
radiation near the field galaxy in consideration.
In Fig.  \ref{fig:frac2} (upper panel) we plot the fraction of externally reionized
field haloes as a function of the surrounding overdensity in collapsed
haloes with mass $M> 5 \times 10^{12} \Msunh$ at $z=0$. The overdensity is
measured within a sphere of radius $10\ h^{-1}\rm Mpc$ centered on the
halo in consideration, does not include this central halo, and is measured relative to
the average density of the simulation box. Contrary to what one might expect, there is no clear
trend. 

This result persists if we change the mass limit on the
surrounding collapsed haloes, or the radius of the sphere. In
Fig. \ref{fig:frac2}, bottom panel, we show for the example of the
WMAP3-f250C simulation that the constant
behaviour of the externally reionized halo fraction also persists if we
divide the halo sample into different mass bins.  As evident in
Fig.~\ref{fig:tim2}, the reason for the constant behaviour of the
external fraction with halo mass is that not only the
reionization, but also the formation of a halo occurs at
slightly lower redshifts in under-dense regions, since the local ``clock'' of structure
formation is slower in voids. 
Therefore, we conclude that the environment of haloes does not change the way in
which they are reionized, at least if the dark matter halo is
isolated. For cluster and group galaxies being part of a larger dark
matter halo, this statement might not be correct anymore.
Satellite galaxies in massive clusters that are close to the center cD galaxy
will most probably be reionized by the time when 70 percent of the dark 
matter mass surrounding the central cD galaxy itself is reionized. This is
clearly earlier than the typical redshift where galaxies in the field
are reionized. However, the difference is probably less dramatic than
the difference in formation redshift. One could speculate that this
means that galaxies in clusters are reionized internally 
more often than galaxies in the field. We have found this to be the
case at least for the resolvable subhaloes in clusters, which we have
identified with our subhalo finder SKID \citep{2001PhDT........21S}. However, since they are not a fair tracer of the
subhalo population, we do not discuss our results in detail.

\subsection{The Reionization History of the Local Group}
\label{LGr}
\begin{figure}
\centerline{\psfig{figure=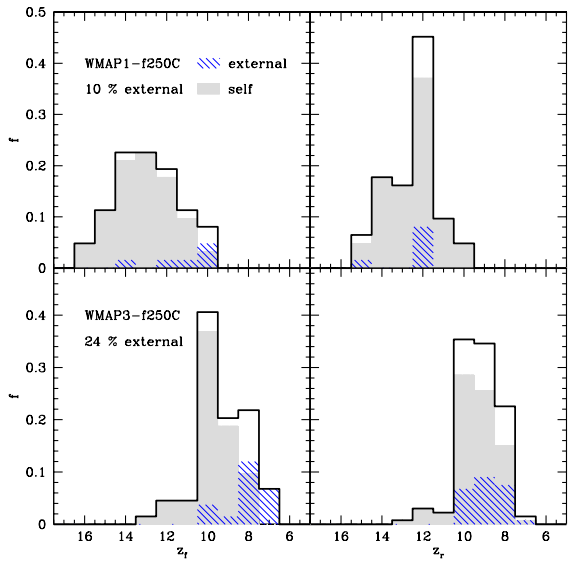,width=3.2in}}
\caption{Comparison between formation redshift ($z_f$) and
  reionization redshift ($z_r$) for LGs that are externally (diagonal texture)
  and internally (grey area) reionized. Results are shown for the
  WMAP1-f250C and the WMAP3-f250C simulation. Again, $z_f$ is defined
  as the redshift where the mass of the halo exceeds $M_{th}$.
}
\label{fig:zreion}
\end{figure}

In this section, we study the reionization history of Local Group like
objects. As binary systems, Local Group like objects might be expected
to have a different average formation
and reionization history than typical field galaxies. Our Local Group of galaxies also provides the most
detailed set of observational data for testing the consequences of
reionization on subsequent galaxy formation.We apply the same definitions  of internal and external reionization we
described above to our LG candidate sample.
For WMAP3-f250C we find that 24\% (32 out of 133 LGs)
are externally reionized. For WMAP1-f250C, the probability of external
reionization is only $\sim 10 \%$ (6 out of 62 LG). 
As for L* and field galaxies, WMAP1-f2000 shows the highest fraction of 
external reionized LGs: 44\% (27 out of 62). In Fig. \ref{fig:zreion}, we show the distribution of the
reionization and formation redshifts of LG candidates 
for the two simulations with the lower reionization efficiency: WMAP1-f250C and WMAP3-f250C.
As in Fig. \ref{fig:zav} at the lower mass end, it is evident that
the scatter in formation redshift is larger than the scatter in
reionization redshift. This is caused by the contribution of
external reionization. 

To summarize, the probability for LG-like objects to be externally reionized 
is lower than the corresponding probability for low-mass field galaxies, 
which can be as high as 60-80\%. 
Compared to typical L* galaxies, LG-like systems form slightly
earlier, get reionized slightly earlier and are somewhat less likely to be
externally reionized than L* galaxies.

\section{Potential Caveats}
\label{sec:caveats}

All of our simulations resolve only haloes with masses above 
$M_{th}\sim 2\times 10^9 M_\odot$ and these are the only sources which
contribute to the reionization of the universe in this set of simulations.
However, in hierarchical structure formation scenarios the smallest haloes 
form first and merge up. Thus it is expected that the first sites of star
formation were ``minihaloes'' with a mass roughly between $10^4$ and 
$10^8  \Msunh$. By definition the minihaloes have virial temperatures below 
$10^4$~K, where the radiative gas cooling through hydrogen 
and helium atomic lines is not efficient. Thus, these haloes have to rely 
on the much less efficient molecular line cooling for their star formation.
However, the hydrogen molecules are easily destroyed by radiative feedback
\citep*[e.g.][]{2000ApJ...534...11H} and thus it is generally expected that 
most minihaloes were `sterilized' and in most cases did not form stars  
\citep[e.g.][and references therein]{2004MNRAS.348..753S,2005MNRAS...361..405I}.
Haloes with a mass between $10^8$ and $10^9 M_\odot$, on the other hand, which 
are also not resolved in our simulations, are able to cool efficiently and 
may actually be an important source of ionizing radiation. \citet{selfregulated}
have performed a number of smaller-scale high-resolution simulations of 
reionization, with a simulation volume $(35\,\rm h^{-1}Mpc)^3$, which include 
radiation from these smaller haloes. The most important effect is that 
reionization starts much earlier than without those sources. However, due to 
Jeans-mass filtering in the ionized regions and the strong clustering of haloes 
around the density peaks, the majority of these low-mass haloes are suppressed 
throughout most of the reionization process and hence their effect is mostly 
limited to the early stages of reionization and the progress of reionization 
is delayed. The time of overlap and the large-scale geometry of reionization 
are both largely dominated by the high-mass haloes and thus are not strongly 
affected by the absence of low-mass haloes in the simulations. 

Could the inclusion of low-mass haloes and minihaloes significantly modify 
our results? A halo with a mass of $10^8 M_\odot$ can reionize a mass between 
$2.5\times 10^{10} - 2.5 \times 10^{11} M_\odot$ (somewhat decreased once 
recombinations are accounted for) during its lifetime of $\sim$20~Myr (before 
star formation shuts off again due to feedback, and recombinations set in, 
decreasing the reionized fraction further). Thus, in order to reionize a 
(proto-)halo with a mass around $10^{12}M_\odot$, several $10^8 M_\odot$ 
haloes ought to have formed at very similar times within the region considered. They 
should also have been spaced far enough apart so that the halo forming first 
does not suppress star formation in the other haloes. Even if this were the case, 
this group of $\sim$10$^8 M_\odot$ haloes would most probably quickly merge, 
forming a halo with a mass above $10^9 M_\odot$, which would be resolved 
in our simulations. Moreover, regions which contain only $10^8-10^9 M_\odot$ 
haloes, but not any of the more massive ones tend to be near the boundaries 
of the ionized regions, and are thus quickly overrun by the expanding 
I-fronts. Furthermore, both low-mass atomically-cooling haloes and minihaloes 
are strongly clustered around the density peaks, which are the first regions 
to be reionized, as we have shown. Therefore the vast majority of these small 
haloes would be affected by Jeans-mass filtering. As a consequence, haloes 
forming after their vicinity is ionized would not be able to accrete gas, 
while haloes that have formed before their region became ionized would either 
not accrete more gas and thus have their star formation shut off (in the case 
of low-mass sources), or would be photoevaporated (in the case of minihaloes).

For all these reasons, we expect that neither haloes with masses between 
$10^8-10^9 M_\odot$, nor minihaloes would change our average reionization
redshifts significantly and hence would not modify our results for the 
external/internal reionization fractions. Once future cosmological simulations 
become large enough to allow resolving low-mass haloes in a large volume, it 
would be interesting to verify if our assertions here are correct and evaluate
the level of this effect precisely.

Furthermore, the reionization parameters, photon production efficiencies  
and small-scale gas clumping (and to lesser  level, also the background 
cosmology) are still quite uncertain. We have addressed this by comparing three  
reionization scenarios which roughly bracket the expected range of
reionization histories, from the very early reionization given by
f2000\_WMAP1  (high-efficiency  sources,  no sub-grid  clumping,  WMAP1
cosmology), to the  late  overlap, extended reionization scenarios,
f250C\_WMAP1 and f250C\_WMAP3 (low-efficiency sources, sub-grid clumping, 
WMAP1 or WMAP3 cosmology). Most of the reasonable reionization scenarios  
which satisfy the current observational constraints should fall between
these cases.  Further improvement  would require better, more detailed
observational  data which would  constrain the  reionization scenarios
(and thus the reionization parameters) better.

\section{Implications of External vs. Internal Reionization regarding 
Galaxy Properties}
\label{sec:impli}

Reionization can affect the properties of galaxies in many ways,
and the relative timing of reionization across the mass scales 
will leave distinct observable signatures.

\begin{itemize}

\item The mode of galaxy formation may be fundamentally different between
haloes less massive and more massive than $10^{12}M_\odot$. Late forming haloes
that are externally ionized will have hotter, more extended and smoother
gas distributions. The gas accretion into the galaxy will proceed differently,
perhaps providing an explanation for the red-blue galaxy sequence, bulgeless 
disks, lower-baryon fractions in less massive galaxies, etc.

\item Early star formation will be affected, since
stars forming in ionized gas may have different properties than
stars forming in neutral gas. 
According to \citet{2006MNRAS.366..247J}, Pop II.5 could
have formed from primordial gas that is strongly ionized since the 
free-electron-catalyzed formation of molecules enables enhanced gas cooling.  
Similarly, \citet{2005MNRAS.364.1378N} predict the formation of low mass 
stars in ionized gas with very low metallicity. External 
reionization provides an alternative pathway to Pop II.5 star formation.

\item The abundance of Galactic satellites has been reconciled with 
observational data using the idea that reionization suppresses the 
formation of late forming haloes - only the rare peaks that collapse 
before reionization can cool gas to form stars. If the gas in a 
late-forming galaxy is externally reionized it may suppress
the formation of some or even all of its satellite population. 

\item The radial distribution of old stars in the Galactic Halo might reflect
the rarity of the density peaks in which they formed at the epoch of
reionization \citep*[][\citealt{2006MNRAS.368..563M}]{2005MNRAS.364..367D}. 
If stars form only in the rarest peaks, then their final distribution can 
be highly concentrated with a radial profile as steep as $r^{-4}$. Thus, 
externally reionized galaxies may be expected to have steeper diffuse 
stellar halo profiles.

\item Similar to the old stellar halo, globular cluster formation may be 
truncated at reionization, altering the relative abundances and radial 
distributions of the old blue population in galaxies of different masses 
and environments. \citet{2005ApJ...630L..21R} found that the number of 
old blue GCs per galaxy stellar mass increases with increasing mass for 
cluster ellipticals, while it seems to be constant for spiral galaxies 
in the field. They suggest that the reason for this is that more massive 
elliptical galaxies assemble earlier and have more time to form blue GC 
before reionization, assuming that reionization happens at a fixed point 
in time. Although we find that reionization will occur considerably earlier 
in massive elliptical galaxies than for low-mass field galaxies, 
their redshift of formation is shifted even more strongly, as 
they are usually internally reionized. Thus it is indeed the case that 
typically more time elapses between the formation and reionization of 
massive cluster ellipticals than for field galaxies.

\end{itemize}

We conclude that it may already be possible to discriminate between 
galaxies that have been reionized internally or externally using one 
or more of the above characteristics. This will be an interesting new
and complementary way to probe the history of reionization in the 
Universe and its effect on galaxy formation.

\section{Summary and Discussion}
\label{sec:resdis}

The reionization history of a given galaxy is strongly dependent on
its mass due to the correlation between galaxy mass and formation
redshift. Higher mass haloes on average form earlier and are also
reionized earlier. The lower the mass of a halo, the more important
does external reionization (i.e. reionization emanating from sources
outside the halo) become. The effect of the large-scale environment
on the reionization history, on the other hand, is negligible for
field haloes.
 By using a
simple polynomial fit, we can parameterize the average reionization redshift
of a galaxy as a function of halo mass for three different
combinations of cosmology and reionizing photon production
efficiency. Typical L* galaxies and LG like systems are reaching
reionization considerably before the final overlap, at a redshift where
roughly half of the Universe is reionized. Very massive, elliptical
galaxies in clusters are reionized even earlier and are always 
internally-reionized. Our prescription for the reionization redshift as 
a function of galaxy mass might be of use for semi-analytical models of 
galaxy formation. Normally, these models assume a fixed, fairly low redshift
of reionization of around $z \sim$7 \citep[e.g.][]{2006MNRAS.365...11C},  
from which the baryon fraction in a low mass haloes can be calculated, 
given redshift and halo mass 
\citep*[\citealt{2000ApJ...542..535G},][]{tumultuous}. Using our fitting 
formulae, one could utilize a more realistic reionization redshifts in 
semi-analytical models by introducing a dependence on the mass of the 
halo in which a given subhalo ends up. This may help alleviate the problem 
of the overproduction of satellite galaxies in massive clusters in 
semi-analytical models \citep{2006MNRAS.372.1161W}.

We suggest that whether a galaxy experiences internal or external
reionization might affect its properties considerably. Externally
reionized haloes could experience different gas accretion;
could have a different first generation of stars (i.e. Pop II.5 instead of
Pop III); may host a lower number of satellite galaxies and blue
globular clusters; and could have steeper diffuse stellar halo profiles.

\section*{Acknowledgments}
Part of the N-body simulations have been performed on the zBox2
supercomputer at the University of Zurich
(http://www-theorie.physik.unizh.ch/$\sim$dpotter/zbox2/).
We thank D. Potter and J. Stadel for building and maintaining
the zBox2. We also thank F. van den Bosch and M. Volonteri for useful 
discussion and an anonymous referee for comments which helped us improve 
the paper. SW has been partially supported by the Swiss National Science 
Foundation (SNF).

\bibliographystyle{mn2e}

\end{document}